\documentclass[aps,preprint]{revtex4}

\def\be{\begin{eqnarray}}
\def\ee{\end{eqnarray}}

\def\l{\lambda}

\def\r2{\sqrt{2}}
\def\rt{\sqrt{3}}
\def\rs{\sqrt{6}}

\def\G{{\cal G}}

\def\({\left(}
\def\){\right)}
\def\f{\phi}
\def\bk#1{\langle#1\rangle}
\def\o{\omega}
\def\1{{\underbar 1}}
\def\p1{{\underbar{1}$'$}}
\def\pp1{{\underbar{1}$''$}}
\def\3{\underbar{3}}
\def\G{{\cal G}}
\def\x{\times}
\def\bi{\begin{itemize}}
\def\ei{\end{itemize}}

\def\bn{\begin{enumerate}}
\def\en{\end{enumerate}}

\begin{document}
\title{Symmetry of Lepton Mixing}
\author{C.S. Lam}
\address{Department of Physics, McGill University,\\ Montreal, QC, Canada H3A 2T8\\
and\\
Department of Physics and Astronomy, University of British Columbia,  Vancouver, BC, Canada V6T 1Z1 \\
\email: Lam@physics.McGill.ca}

\begin{abstract}
Neutrino mixing is studied from a symmetry perspective, both bottom-up and top-down. In the bottom-up approach,
we start from the tri-bimaximal mixing, or one of its three partial patterns, and construct a list of horizontal 
symmetry groups capable of reproducing the mixng without adjustment of parameters. This list,
labeled by an integer $n\ge 3$, is explicitly calculated for $n=3$. In the top-down approach, 
we start from any finite group possessing
a three-dimensional irreducible representation and an order-2 element, give a recipe to determine what mixing
pattern it contains, and how to construct a dynamical model to reveal a particular mixing.  Finally, we 
point out that if 
quark mixing is controlled by symmetry in this way, then
there is an exciting 
possibility to determine most of the CKM mixing parameters by symmetry alone.  

\end{abstract}
\maketitle

\section{Introduction}
Neutrino mixing can be described by the tri-bimaximal PMNS matrix \cite{HPS}
\be
U={1\over\rs}\pmatrix{2&\r2&0\cr -1&\r2&\rt\cr -1&\r2&-\rt\cr}.
\ee
It predicts a zero reactor angle $\theta_{13}$, a maximal atmospheric angle with 
$\sin^2\theta_{23}=0.50$,
and a solar angle with $\sin^2\theta_{12}=0.333$. These predictions agree 
with the measured values $\sin^2\theta_{13}=0.9{+2.3 \atop -0.9}\times 10^{-2}$, 
 $\sin^2\theta_{23}=0.44(1 {+0.41\atop -0.22})$,  and $\sin^2\theta_{12}=0.314(1 {+0.18\atop -0.15})$ deduced 
from a global fit of the experimental data \cite{FLMP} to better than one standard deviation.

There are many attempts to produce an acceptable mixing from a horizontal group
 \cite{S3,A4,S4,SU3,OTHERS}.  A successful model should 
yield the mixing pattern (1)  automatically (or something equal to it within experimental errors) without having
to tune any parameter. 
Parameters present should only be used to fit the neutrino mass gaps but not the mixing pattern.
A partially successful model would be one that reproduces some but not all the features of (1), in which case the 
parameters have to be used to tune the mass gaps as well as the remaining features of mixing.
  With these criteria in mind, the 
degree of success of the existing
models varies, but even when they are (partially) successful, it is not always clear what 
is responsible for the success because there
are so many adjustable inputs.
 Is it the choice of a particular
horizontal group, the choice of particular irreducible representations, and/or  the choice of certain Higgs
expectation patterns? Sometimes different choices can also lead to very similar results, and why is that so?
There is  also  the important but difficult problem of understanding the significance of the mixing pattern (1).
Why is it that way but not something else completely arbitrary?

In this work we attempt to answer some of these questions in a bottom-up approach, in which both (1) and
 three of its partial patterns are studied. The latter are 
bimaximal mixing without trimaximal mixing, tirmaximal mixing without bimaximal mixing,
and a third pattern specified by the first column of $U$ in (1), just like
bimaximal and trimaximal 
mixing are respectively specified by the third and the second columns. We shall label these 
patterns by an index $\alpha$: the full pattern is $\alpha=0$, and the partial pattern 
specified by column $i$ of (1) is $\alpha=i$. For a partial pattern, one column of the mixing
matrix is fixed, and the other two are arbitrary, subject only to unitarity considerations.

For a given $\alpha$, the first question to ask is what finite horizontal groups 
$\G^\alpha$ can automatically give rise to that mixing pattern. A method to construct a 
list is discussed in the next section. The resulting group, 
labeled by an integer $n\ge 3$, will be denoted $\G_n^\alpha$.  Some $(\alpha,n)$ may yield no group, while
others may give rise to more than one. The list may turn out to be finite or
infinite.
For $n=3$, the list is
\be
\G_3^0&=&\G_3^1=\{S_4, H(12,3)\},\quad \G_3^2=\{A_4\},\quad \G_3^3=\{S_3, H(6,3)\},\ee
where
$S_k$ is the symmetric (permutation) group of $k$ objects,
and $A_k$ is the corresponding alternating (even permutation) group. The group $H(m,n)$ is
 defined by two generators $F$ and $G$ with the
relations $G^2=F^n=(FG)^m=E$ (identity matrix) \cite{REL}. In particular, it is known that \cite{PRESENTATION}
$H(2,2)=Z_2\x Z_2,\ H(4,2)=D_4$,\ $H(2,3)=S_3,\ H(3,3)=A_4$, and $H(4,3)=S_4$. The notations $Z_m$ and $D_m$
stand for the cyclic and the dihedral group respectively. 
The group $H(6,3)$ has 54 elements, and the group $H(12,3)$ has 216 elements. Clearly any group that contains a
subgroup in this list will do as well. 
For example, $A_4\in\G_3^2$ can guarantee a trimaximal mixing ($\alpha=2$) but not
a bimaximal mixing ($\alpha=3$). However, since $S_3,A_4\in S_4$, the group $S_4$ can produce not only the mixing pattern
$\alpha=1$, but also trimaximal and bimaximal mixing, and that is why $\G_3^1=\G_3^0$.

The next question to ask is what $\G$-representations and Higgs expectation
patterns are needed to yield the specific mixing pattern. This is discussed in Sec.~3, using an approach
in which the right-handed and all heavy leptons have been integrated out, so only left-handed leptons and Higgs remain.
With two exceptions which will be discussed in Sec.~3, 
the left-handed leptons are always assigned a three-dimensional irreducible representation, but it does not
matter how many Higgs are present and what representations they belong to, 
as long as their expectation values are invariant under
the appropriate residual symmetry operators to be discussed in Secs.~2 and 3. For dynamical models with the presence
of right-handed and/or heavy leptons, the present formalism gives only the constraint placed on the effective model after
these other leptons are integrated out.

As to the significance of having the tri-bimaximal mixing pattern
rather than something arbitrary, we can
offer the following observation which will be elaborated in Sec.~3. Any {\it finite} horizontal group
that contains the requisite residual symmetries can only yield a small number of possible mixing patterns; 
the tri-bimaximal pattern (1) and
its sub-patterns are among those possible 
for the groups in $\G_n^\alpha$. In general, the smaller the horizontal group, the more limited is the number of possible
mixing patterns. Furthermore, the choice of horizontal groups is quite limited, as it must be a group containing an
order-2 element to act as the neutrino residual symmetry, and a three-dimensional irreducible representation
for the leptons to reside in. Thus, although
symmetry alone cannot determine mixing, it can pick out a discrete number of possibilities to which (1) and its 
subpatterns belong.

That also means that a mixing pattern
which deviates from (1) by a small amount in all three columns is expected
to come either from an infinite horizontal group, or else the residual symmetries are softly but weakly broken.

It would be very exciting if quark mixing can be explained by symmetry in the same way, for the discreteness of the 
allowed mixings makes it conceivable to have most of the CKM parameters determined 
that way by symmetry alone. This point
is further discussed in the concluding section.

\section{Residual and Horizontal Symmetries}
Since symmetries normally refer to Hamiltonians, in this case mass matrices, it is useful to find out how to translate 
mixing patterns into mass-matrix symmetries. In the basis of a diagonal charged-lepton mass matrix (squared) 
$M_eM_e^\dagger$, the Majorana mass matrix $M_\nu$ for the active neutrinos can be diagonalized by the mixing matrix $U$ 
to produce
$U^TM_\nu U={\rm diag}(m_1,m_2,m_3)$. Using this formula, it can be shown that bimaximal mixing 
is equivalent to a 2-3 symmetry
of the
mass matrix $M_\nu$ \cite{23}, and trimaximal mixing is equivalent to a magic symmetry \cite{MAGIC}. The
2-3 symmetry defined by the relations $(M_\nu)_{12}=(M_\nu)_{13}$ and $(M_\nu)_{22}=(M_\nu)_{33}$ 
is generated by a unitary matrix $G_3$ commuting with $M_\nu$, and the magic symmetry
characterized by the magic-square property of having equal sums for rows and columns  
is generated by another unitary matrix $G_2$ commuting with $M_\nu$  \cite{TEXTURE}. 
We shall refer to these symmetries as  {\it residual symmetries}
of $M_\nu$.
The residual symmetry group generated by $G_2$ and $G_3$
is a $Z_2\x Z_2$ group. It  contains the element $G_1=G_2G_3=G_3G_2$ which generates the mixing pattern
of the first column of $U$ in (1). Each $G_i$ generates a subgroup of $Z_2\x Z_2$ 
isomorphic to $Z_2$. 

As a matter of fact, such generators $G_i$ can be constructed for {\it any} real PMNS matrix $U$ as follows. Let $v_i$ denote
the $i$th column of $U$, then the three $v_i$ form an orthonormal set, and the equation $U^TM_\nu U={\rm diag}(m_1,m_2,m_3)$ is equivalent 
to the eigenvalue equations $M_\nu v_i=m_i v_i$. The matrix $G_i=-E+2v_iv_i^\dagger$ has eigenvalue 1 with the eigenvector
$v_i$, and a degenerate eigenvalue $-1$ for the other two eigenvectors. We may therefore choose the other two eigenvectors
of $G_i$ to be $v_j$ and $v_k$, with $i\not=j\not=k\not=i$. Since $G_i$ has the same eigenvectors as $M_\nu$, clearly
the two commute. It is also easy to verify that $G_i$ is unitary, $G_i^2=E$, and $G_iG_j=G_jG_i=G_k$. Hence 
a residual symmetry group $Z_2\x Z_2$ generated by these $G_i$ is present for any real 
PMNS matrix $U$. The difference between
the $G_i$'s of different $U$'s is their explicit matrix form, and not the group structure.
In the case of the tri-bimaximal mixing in (1), the generators can be computed to be \cite{TEXTURE}
\be
G_1={1\over 3}\pmatrix{1&-2&-2\cr -2&-2&1\cr -2&1&-2},\quad
G_2=-{1\over 3}\pmatrix{1&-2&-2\cr -2&1&-2\cr -2&-2&1},\quad
G_3=-\pmatrix{1&0&0\cr 0&0&1\cr 0&1&0}.\ee

\vspace{.3cm}
What about residual symmetries of the charged-lepton mass matrix? Since it is diagonal and non-degenerate, any unitary
matrix $F$ that commutes with it must be diagonal with unit modulus in all its entries. 
We shall assume the presence of at least one residual symmetry
$F\not=E$ with non-degenerate eigenvalues. This simple but powerful assumption allows
us to construct a horizontal symmetry group $\G$ generated by $F$ and one or 
two (the third one is not independent) $G_i$. In other words, $\G^i=\{F,G_i\}$ and $\G^0=\{F,G_2,G_3\}$. 
We shall also
assume the horizontal group to be a finite group.

The reason for requiring $F$ to be non-degenerate is to ensure $M_eM_e^\dagger$, which commutes with $F$,
to be diagonal when $F$ is. This is necessary since the neutrino residual symmetries $G_i$ in (3) are defined in the basis 
where $M_eM_e^\dagger$ is diagonal. This is also the reason why we need to have a residual symmetry 
$F$ for the charged-leptons.

The reason to take $\G$ to be a finite group is economy. Since all the symmetries
in $\G$ other than the residual symmetries are eventually broken, there is
no point to throw away more than necessary by starting with a larger group. With a finite group,
especially a small one, the theory also has more predictive power. 
Given a $F$, suppose the pair $\{F, G_i\}$ generates a finite group. Among other things this means the
existence of an integer $m$ so that $(G_iF)^m=E$.
Clearly this relation can no longer be satisfied if we make a small change of $G_i$, so the resulting new group 
is either
infinite or very large. See Sec.~3 for further discussions on this point.
  Hence finite-group requirement has the
power to limit $G_i$ to a discrete number of choices. 

Since $\G$ is a finite group, there must be an integer $n$ for which
 $F^n=E$. The corresponding group is labeled $\G^\alpha_n$.
Given an $n$, the three matrix elements of the diagonal $F$ must each be an $n$th root of unity,
hence there are
$n(n-1)(n-2)$ choices of $F$ with non-degenerate eigenvalues. In particular we need to have $n\ge 3$.
Since every $G_i$ in (3) is 2-3 symmetric, it will
not produce anything new if we interchange the 2,\ 3 elements of $F$, hence there are only $n(n-1)(n-2)/2$ distinct ones.
If we label them by an index $a$, and the corresponding diagonal matrix denoted as $F_{na}$, then the group
generated by $\{F_{na},G_i\}$ or $\{F_{na},G_2,G_3\}$ may or may not be finite. If it is not finite, then we reject
it and go on to another $F$. If it is finite, we include it in the list for $\G_n^\alpha$.

There is the practical matter of deciding whether the group generated by $\{F,G\}$ is finite or not. 
A necessary condition is that $GF$ must have a finite order $m$: $(GF)^m=E$. If necessary we can resort to numerical 
means by computing the eigenvalues of $GF$. For $m$ to be finite, every one of the eigenvalues must have modulus one,
 and its phase angle divided by
$2\pi$ must be a rational number. When computed numerically, every floating point number can be approximated by a rational
number, but if the denominator of this rational number is larger than 1000, I shall declare the number to be irrational
and that $F$ rejected. 
Otherwise we know what $m$ is. The resulting group, generated by $\{F,G\}$
with the relations $G^2=F^n=(GF)^m=E$, will be denoted as $H(m,n)$ \cite{REL}. For some $m$ and $n$ such a group is explicitly
known \cite{PRESENTATION}. For others we do not even know whether the group is finite, in which case we must compute
the multiplication table to find out.

For $n=3$, there are three distinct $F$'s: $F_{31}={\rm diag}(1,\omega,\omega^2), F_{32}={\rm diag}(\omega,1,\omega^2),
F_{33}={\rm diag}(\omega^2,1,\omega)$, where $\omega=\exp(2\pi i/3)$.
 The groups produced by $F_{31}$ are $\G^1_3=S_4,\ \G^2_3=A_4,\ \G^3_3=S_3$, and
$\G^0_3=S_4$, The groups produced by $F_{32}$ and $F_{33}$ are identical and they are 
$\G^1_3=\G^0_3=H(12,3),\ \G^2_3=A_4,$ and $ \G^3_3=H(6,3)$. The group $\G^2_3=A_4$ has been discussed
in \cite{AF} in a similar way.

For $n=1$ and $n=2$, the 3$\x$3 matrix $F$ cannot be non-degenerate. If we ignore this requirement, then
we already know the $n=1$ result
$\G_1^i=Z_2$ and $ \G^0_1=Z_2\x Z_2$. For $n=2$, $\G^{0,1,2}_2$ do not exist and $\G^3_3=\{Z_2\x Z_2, D_4\}$.
These are perfectly legitimate
horizontal groups but since $F$ has degenerate eigenvalues,
the mixing patter $\alpha$ cannot be {\it automatically} produced.

It should be emphasized that although $\G_n^\alpha$ is computed when $M_eM_e^\dagger$ is diagonal, nothing really depends
on it. In any other basis, the mass matrices undergo a unitary transformation, $M\to V^\dagger MV$, but they still
commute with the transformed residual symmetry operators $V^\dagger(F,G_i)V$. Moreover, the structure of $\G$ remains the same
with the transformed group elements.

\section{Dynamical Models}
Once a horizontal group in $\G_n^\alpha$ is chosen, a dynamical model with the mixing pattern $\alpha$ 
can be produced simply by breaking all but the residue symmetries, either softly or spontaneously. 
Before going into the specifics there, let us 
first discuss how to produce a mixing pattern from
{\it any} finite group $\G$, not necessarily one that is known to be in some $\G_n^\alpha$.

We consider mass terms only after all right-handed fermions and heavy leptons are integrated out, in which case
only the left-handed leptons $L=(e,\nu)$ and the Higgs are left. 
With two exceptions to be discussed below, 
the left-handed fermions are assigned
to a three-dimensional irreducible representation (3D IR) of $\G$. For dynamical models with the presence
of right-handed and/or heavy leptons, the present formalism gives only the constraint placed on the effective model after
these other leptons are integrated out.

There are several requirements for $\G$ to satisfy before it is qualified to be a horizontal group. Besides having
a 3D IR for the left-handed fermions to occupy,
it must also have
an order-two
element $G$ to act as the residual symmetry of the neutrino matrix. This calls for groups with an even
order.  $G$ must have one $+1$
eigenvalue with some eigenvector $v$ which defines the partial mixing pattern, and two $-1$ eigenvalues. 

Let us see what happens if $L$ does not belong to a 3D IR. In that case it either contains  
three 1D 
IR or one 2D IR and a 1D IR. In the first situation if the values of $G$ in the three 1D IR are $a,b,c$ respectively,
then $G$ must be the diagonal matrix $G={\rm diag}(a,b,c)$. To have the correct eigenvalues, one of the three
must be $+1$ and the other two $-1$. The eigenvector $v$ of eigenvalue $+1$ then has two zero entries, and the third
one equals to 1. The unitary mixing matrix $U$ containing $v$ in one of its columns must be block-diagonal, so only
two of three leptons can mix. In the second situation $G$ must be block diagonal itself, with either $+1$ or $-1$
appearing in the $1\x 1$ block. If that is $+1$ then $v$ and the mixing matrix $U$ are the same as in the former case.
If it is $-1$ then $v$ must come from the $2\x 2$ block, hence it must have a zero entry like the third column of (1).
This is why the group $S_3$ appears in $\G_3^3$ of (2) and nowhere else, because $S_3$ only has a 2D IR but not a 3D IR.
Besides these two exceptional cases, $L$ must belong a 3D IR, as claimed.

Each of the charged-lepton mass (squared) term is of the form $C_{ija}e^\dagger_i\phi_a^e e_j$, and 
each of the Majorana neutrino
mass term is of the form $C_{ija}\nu^T_i\phi_a^\nu \nu_j$, where $\phi_a^{e,\nu}$
are the Higgs fields, and $C_{ija}$ are the Clebsch-Gordan coefficients. The $\G$-structure is 
explicit in this notation but all the Standard Model structures are understood and ignored.
There may be many such mass terms, each
with a different Higgs, a different Clebsch-Gordan coefficient, and a different Yukawa coupling constant.
We have assumed the couplings to be linear in the Higgs; otherwise we simply consider $\f^{e,\nu}$ to be composite fields.
The representation of the Higgs can be either reducible or irreducible, as long as each of the mass terms is invariant
under $\G$.

Under a transformation induced by $g\in\G$, $\nu\to g'\nu,\ e\to g'e,\ \f_a\to g''\f_a$, where $g'$ is a 3D IR
 of $g$, and $g''$ is the appropriate representation for the $\G$-multiplet $\phi$, which can be either reducible
or irreducible. The invariance
of the mass terms demands $C_{ija}=g'_{ik}g'_{jl}g''_{ab}C_{klb}$ for the neutrinos and 
$C_{ija}={g'}^*_{ik}g'_{jl}g''_{ab}C_{klb}$ for the charged-leptons. 

The introduction of vacuum expectation values $\bk{\phi}$ breaks $\G$ down to the residue symmetries.
The residual symmetry of neutrinos is by construction $G$, but we must 
still decide on the residue symmetry $F$. In principle,
it can be any element of $\G$ as long $F'$ has three distinct eigenvalues. If its order is $n$, then we must have
$n\ge 3$ for $F$ to have all distinct eigenvalues.

The mass matrices $(M_eM_e^\dagger)_{ij}$ and $(M_\nu)_{ij}$ are of the form $C_{ija}\bk{\phi_a}$.
A sufficient condition for these mass terms to be invariant under the residue symmetries $F$ and $G$ is 
$F''\bk{\phi^e}=\bk{\phi^e}$ and $G''\bk{\phi^\nu}=\bk{\phi^\nu}$. Now go to the basis where $F'$ is diagonal.
Since its eigenvalues are all distinct and it commutes with $M_eM_e^\dagger$, the charged-lepton mass matrix must also
be diagonal in that basis. The eigenvector $v$ of $G'$ with eigenvalue $+1$ then defines a partial mixing pattern, as
it occupies one column of the mixing matrix $U$.

These arguments apply just as well to the special case when $\G=\G_n^\alpha$, except that in this case we simply take
$F'=F_{na}$ and $G'=G_i$ when $\alpha=i$. When $\alpha=0$, we will repeat the construction for two different $i$'s.

If $\bk{\f^e}$ and/or $\bk{\f^\nu}$ is  a $\G$-triplet in the same representation as the leptons, 
or tensors of such triplets,
then much more can be said
because we know then $F''=F'=F_{na}$ and/or $G''=G'=G_i$. 
Since the diagonal
matrix $F_{na}$ has three different entries along its diagonal, this invariance for $\bk{\f^e}$
is possible only when one of the diagonal entries of $F_{na}$ is 1, 
and the components of the $\bk{\phi^e}$ facing the other two entries are zero. Thus dynamical models constructed from 
a horizontal group without an entry 1 in its $F_{na}$ to start with
may not have a triplet Higgs of this kind with non-zero expectation values. 
On the neutrino side, since $G_iv_i=v_i$,
$\bk{\phi}$ must be proportional to this $v_i$. 
This automatically prevents a full symmetry pattern $\alpha=0$
to be obtained 
if triplet Higgs are present because $\bk{\phi}$ cannot be proportional to two different $v_i$'s.

In the remainder of this section, two examples of $\G=A_4$  from the literature will be taken 
to illustrate some
of these points.   

We see from (2) that $A_4$ comes from $\G_3^2$, hence the only residual symmetry we can expect to obtain 
from this group without tuning parameters is the magic symmetry corresponding to trimaximal mixing with
of $i=2$. 
Let us now look at the specific example taken from Ref. \cite{M1}, where the Higgs for charged leptons
(called $\Phi$) is a ($\G$-)triplet with expectation values $\bk{\Phi}=v(1,1,1)^T$, and $M_e$ is not diagonal. 
Note that this is the Higgs coupling the left-handed charged fermions to the right-handed ones, so it can be
related to our Higgs $\f^e$ only after the right-handed leptons have been integrated out. The resulting $\bk{\f^e}$
is the tensor product of two triplets.
On the neutrino side, there is a triplet Higgs field called $\xi_{4,5,6}$
and three \1, \p1, \pp1 Higgs fields called $\xi_1,\xi_2,\xi_3$, with expectation values 
$(\bk{\xi_4^0},0,0)^T$ and $\bk{\xi_1^0}, \bk{\xi_2^0}, \bk{\xi_3^0}$ respectively. 

To compare with the general theory discussed above we must first transform everything to a basis where $M_eM_e^\dagger$
is diagonal. The unitary matrix to do that is
\be
V={1\over \sqrt{3}}\pmatrix{1&1&1\cr 1&\o^2&\o\cr 1&\o&\o^2\cr}.\ee
The transformed expectation values, distinguished by a hat, are $\bk{\hat\Phi}=\sqrt{3}v(1,0,0)^T$
and $\bk{\hat\xi_{4,5,6}}=(\bk{\xi_4^0}/\sqrt{3})(1,1,1)^T$. 
Thus $\bk{\hat\Phi}$
is indeed invariant under $F''=F_{31}={\rm diag}(1,\omega,\omega^2)$, and $\bk{\hat\xi_{4,5,6}}$
is indeed invariant under $G''=G_2$ with eigenvalue 1, as the general theory indicates.
 As for the three singlet representations, we have to know 
the values of $G''$. This is given in Reefs. \cite{AF} and \cite{MR} and 
they are all 1 for \1,\p1, and \pp1. Hence $\bk{\hat\xi_{1,2,3}}$
are invariant under $G''$ as well, as the general theory demands. With all these fulfilled, the mixing pattern should
correspond to a trimaximal mixing ($i=2$), with a magic neutrino mass matrix. This
can be seen to be the case in CEQ.~(19) of Ref. \cite{M1}.

The second example is taken from Ref. \cite{M2}. In this case $M_e$ is diagonal so the general theory
should apply without a transformation. On the charged lepton
side, there are three Higgs belonging to \1, \p1, \pp1 respectively, whose expectation values are
$\bk{\f^e}=v_1, v_2, v_3$. 
Since $F''$[\1, \p1, \pp1]=$[1,\o^2,\o] $ \cite{AF,MR},
$\bk{\phi^e}$ is not invariant under $F''$ but $\bk{\f^e}\bk{\f^e}^*$ is, which is all
that counts for $M_eM_e^\dagger$. 
 On the neutrino side, a type-I seesaw mechanism is invoked. The Higgs which couples
 the light to the heavy neutrinos belongs
to a \3, with expectation values $h\equiv(u_1,u_2,u_3)^T$. After integrating out the heavy neutrino, the effective Higgs 
expectation values $\bk{\f^\nu}$ used
to calculate $M_\nu$ is $h\otimes h$. $h$ is invariant under $G''=G_2$ if
$u_1=u_2=u_3$. In that case $G_2$ is a residual symmetry of $M_\nu$ and the neutrino mass matrix should be
magic. This can be seen to be true from eq.~(5) of Ref. \cite{M2}. On the other hand, if 
$u_1=u_2=u\not=u_1$ is assumed, then $h$ is not invariant under $G_2$, so $G_2$ is no longer a residual symmetry,
and $M_\nu$ is not  expected to be magic. This can also be verified from eq.~(8) of Ref. \cite{M2}.

\section{Conclusion}
We have studied the symmetry of neutrino mixing, assuming one residual symmetry each survives for the charged
leptons and the neutrinos after the spontaneous breakdown from a finite horizontal symmetry group. 
The problem is studied both from the bottom up and from the top down.
In the former approach, starting
from any one of the three partial mixing patterns defined by (1), a list of horizontal groups labeled by $n$
can be constructed to yield a specific pattern, and the result for $n=3$ is explicitly shown in eq.~(3). 
In the latter approach, a necessary condition
for a finite group to be horizontal is given, and a general recipe is provided to decide what mixing patterns
it contains, and how to produce an effective dynamical models
giving rise to such a pattern automatically after the right-handed and the heavy leptons are integrated out.

It is natural to ask whether this approach also works for quark mixing. Since the Cabibbo angle is small, the CKM
matrix is not as neat as (1). In any case, it is given numerically and no analytical approximation such as (1)
is known,  so the finite horizontal groups for it are much harder to find. At this point it is not clear
whether such a finite group exists or not for the mixing of three quarks. If it does, then the discreteness of the
allowed mixing patterns for any horizontal group promises the exciting possibility of determining three CKM
mixing parameters once the fourth 
 of them is used to fix which discrete set it belongs to. To test whether a finite horizontal
group exists for quark mixing, I consider the much simpler situation of mixing only two quarks, in which case the mixing
matrix is controlled by a single parameter $\l$, the sine of the Cabibbo angle. Details will appear elsewhere.
I find that this mixing matrix
can indeed be accommodated by the dihedral group $D_m$, for a discrete set of Cabibbo angles given by the formula $\theta_c=
\pi/2m$. For $m=7$, the value of $\l$ is $\l=0.2225$, 
which is to be compared with the Particle Data Group value of either
$\l=0.2272\pm 0.0010$, or $\l=0.2262\pm 0.0014$, from two different fits. Although the predicted value is not close
enough to the experimental result, it is nevertheless way within $O(\l^2)$ of it, an error which we might expect to make
by ignoring the third quark in the mixing. This result gives some hope for the feasibility of the scheme, but to be sure we need to find at least one finite group which contains the mixing of three quarks. This investigation is underway.


\begin{thebibliography}{9}
\bibitem{REL} The group $(2,n,m)$ generated by $F$ and $G$ with the relation $G^2=F^n=(GF)^m=E$ is known in the
literature as the {\it polyhedral group}.  It is finite if and only if $2(n+m)>mn$ \cite{PRESENTATION}. Thus the
finite polyhedral groups are $(2,2,n), (2,3,3), (2,3,4),$ and $(2,3,5)$. For these $n$ and $m$, $H(m,n)=(2,n,m)$.
Otherwise, $H(m,n)$ is a finite subgroup of $(2,n,m)$ obtained by the explicit $3\times 3$ matrices $F$ and $G$
in the text.
\end{thebibliography}
\end{document}